\documentclass[12pt]{article}
\usepackage[utf8]{inputenc}
\usepackage{amsfonts,epsfig}
\usepackage[hyphens]{url}
\usepackage{hyperref}
\usepackage{breakurl}
\usepackage[authoryear]{natbib}

\usepackage{geometry} 
\usepackage{sectsty} 
\usepackage[normalem]{ulem} 
\sectionfont{\sffamily\bfseries\upshape\large}
\subsectionfont{\sffamily\bfseries\upshape\normalsize} 
\subsubsectionfont{\sffamily\bfseries\upshape\normalsize} 
\makeatletter
\usepackage{graphicx}
\usepackage{wrapfig}
 \usepackage{graphicx,fullpage,amsmath,multicol,multirow,amssymb,amsbsy,pifont}
 \usepackage{setspace}
 \usepackage{epsfig}
 \usepackage{amsmath, amsthm, amssymb, bm}
 \usepackage{color}
\usepackage{graphics}
\usepackage{booktabs}
\usepackage{makecell}
\usepackage{caption,array}
\newcommand{\rowgroup}[1]{\hspace{-1em}#1}

\newtheorem{corollary}{Corollary}

\renewcommand\@seccntformat[1]{\csname the#1\endcsname.\quad}
\newcommand{\indep}{\perp \!\!\! \perp}

\makeatletter
\def\@maketitle{%
 \begin{center}%
 \let \footnote \thanks
  {\large \@title \par}%
  {\normalsize
   \begin{tabular}[t]{c}%
    \@author
   \end{tabular}\par}%
  {\small \@date}%
 \end{center}%
}
\makeatother

\title{\bf On the Use of Auxiliary Variables in Multilevel Regression and Poststratification}

\author{Yajuan Si\footnote{University of Michigan; email: \url{yajuan@umich.edu}}}
\date{March 2, 2023}

\begin{document}
\maketitle
\begin{abstract}

Multilevel regression and poststratification (MRP) is a popular method for addressing selection bias in subgroup estimation, with broad applications across fields from social sciences to public health. In this paper, we examine the inferential validity of MRP in finite populations, exploring the impact of poststratification and model specification. The success of MRP relies heavily on the availability of auxiliary information that is strongly related to the outcome. To enhance the fitting performance of the outcome model, we recommend modeling the inclusion probabilities conditionally on auxiliary variables and incorporating flexible functions of estimated inclusion probabilities as predictors in the mean structure. We present a statistical data integration framework that offers robust inferences for probability and nonprobability surveys, addressing various challenges in practical applications. Our simulation studies indicate the statistical validity of MRP, which involves a tradeoff between bias and variance, with greater benefits for subgroup estimates with small sample sizes, compared to alternative methods. We have applied our methods to the Adolescent Brain Cognitive Development (ABCD) Study, which collected information on children across 21 geographic locations in the U.S. to provide national representation, but is subject to selection bias as a nonprobability sample. We focus on the cognition measure of diverse groups of children in the ABCD study and show that the use of auxiliary variables affects the findings on cognitive performance.


\end{abstract}

\section{Introduction}

Nonprobability samples are increasingly common due to declining response rates and rising costs of probability sampling. These samples offer detailed outcomes with large sample sizes, but selection mechanisms may be voluntary or deterministic, and the inclusion (selection and response) probabilities are unknown. The Adolescent Brain Cognitive Development (ABCD) study is a nonprobability sample aimed at national representation but subject to selection and nonresponse bias~\citep{ABCDwebsite}. The study has enrolled participants from 21 U.S. research sites, with convenience and operational constraints dictating the selection process, and the sample enrollment is conditional on the school and parental consent. Lack of randomization and sampling frames undermines the design-based survey inference framework, making statistical validity dependent on the quality of auxiliary information and population model specification~\citep{Smith_1983}. 

Current approaches for inferences with nonprobability surveys rely on calibration with either a reference probability sample or population control information~\citep{elliott-valliant17}. The design-based approach assumes quasi-randomization and constructs pseudo-weights based on the concatenated probability and nonprobability samples. Model-based approaches fit a model of the survey outcome based on the nonprobability sample and predict the outcome for the nonsampled population units~\citep{ghosh:meeden:97,kimetal21}. Doubly robust (DR) methods combine the two approaches and propose a weighted estimator that is also a function of predicted outcomes to offer protection against model misspecification for the sample inclusion probability or the outcome~\citep{wu2018,Yang19}. \cite{kang:schafer07} find that many DR methods perform better than simple inverse-probability weighting; however, none of those studied has improved upon the performance of simple regression-based prediction of the nonsampled values.

As a prediction approach, multilevel regression and poststratification (MRP, \cite{gelman:little-97}) has become increasingly popular. Originally applied to estimate state-level public opinions from sociodemographic subgroups using sample surveys, MRP has two key components: 1) multilevel regression for small area estimation by setting up a predictive model with a large number of covariates and regularizing with Bayesian prior specifications, and 2) poststratification to adjust for selection bias and correct for imbalances in the sample composition. The flexible modeling of survey outcomes can capture complex data structures conditional on poststratification cells, which are determined by the cross-tabulation of categorical auxiliary variables and calibrate the sample discrepancy with population control information~\citep{holt:smith:79,gelman:carlin:00}. 

Besides applications in social sciences, especially in the U.S. and U.K. election forecasting (e.g.,~\cite{lax:philips:09a,lax:philips:09b,wang:gelman14,Yougov:mrp20,MRP:Zahorski20,Yougov2020}), MRP has also shown promise in public health research (e.g., \cite{Zhang15-mrp,Carlin:AJE2018,carlin:AJE20}). For example, MRP has been actively used for COVID-19 viral transmission and immunity prevalence tracking~\citep{specificity:gelman20,mrp-covid21,mrp-covid22}. We aim to apply MRP to the ABCD study, a large-scale survey designed to assess the target population of the U.S. 9 or 10-year-old children with diverse biological, familial, social, and environmental factors~\citep{ABCDdesign}. By calibrating the ABCD sample with auxiliary information about the population, MRP has the potential to yield valid estimates with generalizability, particularly for minority groups with small sample sizes.

However, criticisms of MRP emerge, noting substantial variation in the performances~\citep{buttice_highton,Valliant19-jssam}. \cite{Baker_2013} emphasize that inference for any probability or nonprobability survey requires reliance on modeling assumptions, a coherent framework, and an accompanying set of measures for evaluating the quality. \cite{Yougov:mrp20} consider three polling applications and conclude that a careful model specification is essential in applying MRP. As a statistical method, MRP could be affected by poorly predictive auxiliary information, model misspecification, and invalid assumptions. Nevertheless, there is a lack of guidance on selecting and modeling auxiliary information in practical applications of MRP.  This paper fills in the knowledge gap by examining the finite population inferential validity of MRP and proposing a new data integration framework for robust survey inferences. 

MRP, deeply rooted in survey methodology, uses poststratification to calibrate sample data with auxiliary information. MRP accounts for design features and response mechanisms in prediction modeling, similar to the model-assisted estimation~\citep{sarn:wret:1992,model-assist-review-SS17}. Existing MRP approaches integrate the sample data with auxiliary information from census records or large-scale survey data with low variability, e.g., the American Community Survey (ACS) or the Current Population Survey (CPS). The availability and quality of external information and the integrating method affect the inferential validity. As post-adjustments after data collection, MRP can unify inferences for probability and nonprobability samples under a data integration framework. The probability sample can infer population distributions and generate synthetic populations but has a large sampling error, especially with small sample sizes. Assuming quasi-randomization of the inclusion mechanism into the synthetic population, we estimate the inclusion probabilities for the nonprobability sample units. We account for the uncertainty in synthetic population generation, include flexible functions of estimated inclusion probabilities as predictors in the outcome modeling, and generalize MRP as a unified framework for data integration and robust survey inferences to cope with challenging data settings in practice.

The paper structure is organized as below. In Section~\ref{method} we describe MRP methods on the use of auxiliary information and propose a new framework for data integration. We use simulation studies to compare existing MRP, the proposed framework, and alternative adjustment methods in Section~\ref{simulation} and demonstrate the application to the ABCD nonprobability survey in Section~\ref{application}. We summarize existing challenges and potential extensions in Section~\ref{discussion}.

\section{Methodology}
\label{method}

\cite{rubin83-pi} and \cite{little83-pi} argue that any model for survey outcomes should be conditional on all the information that predicts inclusion probabilities. Suppose the outcome in the population is $Y_i$, the inclusion indicator is $I_i$, and the auxiliary variables of the population are denoted by $X_i$, for $i=1,\dots, N$, where $N$ is the population size. We consider the inference framework~\citep{Smith_1983}
\[f(Y_i, I_i|X_i)=f(Y_i|I_i,X_i)f(I_i|X_i),\]
the validity of which relies on the inclusion mechanism $f(I_i|X_i)$ and the outcome model specification $f(Y_i|I_i,X_i)$. When the outcome $Y_i$ is correlated to the inclusion indicator $I_i$, the inclusion mechanism is informative and has to be accounted for in the analysis of sampled data $f(Y_i|I_i=1, X_i)$. In an ideal design with a simple random sample (SRS) or a well-controlled probability sample with fixed selection probabilities, the selection mechanism can be missing completely at random (MCAR); however, nonrepresentativeness is still a problem for the final sample since nonresponse is inevitable and low response rates often occur. Thus, the inclusion mechanism will deviate from MCAR. The final sample could deviate from the population by unequal probabilities of unit selection, nonresponse, and coverage error. Survey practice often assumes that the inclusion mechanism is ignorable, i.e., missing at random (MAR), given the auxiliary information, $f(Y_i|I_i,X_i)=f(Y_i|X_i)$~\citep{rubin:1976}. However, this requires a correct outcome model specification with rich, highly predictive information $X_i$ from integrated data sources. Probability surveys gather auxiliary information from the sampling frame and linked administrative data or census records. With nonprobability samples, we treat the inclusion mechanism as a quasi-randomized probability sample design and demand auxiliary information about the target population for inferences. 

By constructing poststratification cells with discretized variables $X_i$, we elaborate MRP methods on the use of such categorical auxiliary information to achieve inferential validity and balance estimation bias and variance. We first connect MRP with the literature on poststratification and robust survey inferences and then present the proposed framework for data integration. We focus on the descriptive summaries of the population: the overall mean and subdomain inferences, as these are popular quantities of interest in survey inference.

\subsection{Poststratification}
\label{method-ps}
We start by estimating the population mean of a single survey outcome: $\overline{Y}=\frac{1}{N}\sum_{i=1}^NY_i$. Assume the population means in the respondents and nonrespondents are $\bar{Y}_R$ and $\bar{Y}_M$, respectively. The population proportion of respondents is $\psi$. The overall population $\bar{Y}$ is given by
$\bar{Y}=\psi\bar{Y}_R+(1-\psi)\bar{Y}_M$. Suppose units in the population and the sample can be divided into $J$ mutually exclusive and inclusive poststratification cells with population cell size $N_j$ and sample cell size $n_j$ for each cell $j=1,\dots, J$, with the population size $N=\sum_{j=1}^JN_j$ and the sample size $n=\sum_{j=1}^Jn_j$. Let $\overline{Y}_j$ be the population mean and $\bar{y}_j$ be the sample mean within cell $j$. The overall mean in the population is
$
	\overline{Y}=\sum_{j=1}^J\frac{N_j}{N}\overline{Y}_j.
$
For subdomain estimation, considering poststratification cells constructed at the finest level, we will need to group the set of cell-wise estimates that belong to the subdomain. Let the poststratification cell means for respondents and nonrespondents be $\bar{Y}_{jR}$ and $\bar{Y}_{jM}$, respectively, and the population proportions of respondents in cell $j$ be $\psi_{j}$. The population mean can be expressed as
\[
\bar{Y}=\sum_{j=1}^J\frac{N_j}{N}\left(\psi_{j}\bar{Y}_{jR}+(1-\psi_{j})\bar{Y}_{jM}\right).
\]
To make inference about $\bar{Y}$ with the sample data, the sample mean is the unweighted (UnW) estimator,
\begin{align}
\label{uw}
\bar{y}_{s}=\sum_{j=1}^{J}\frac{n_j}{n}\bar{y}_j.
\end{align}
The poststratification (PS) estimator accounts for the population cell sizes as a weighted average of sample cell means,
\begin{align}
\label{ps}
\bar{y}_{ps}=\sum_{j=1}^{J}\frac{N_j}{N}\bar{y}_j.
\end{align}

Motivated by the model-based perspective of poststratification~\citep{Fay:herriot1979,little93}, \cite{gelman:little-97} propose MRP by modeling cell estimates and predicting nonsampled cases. Assuming that the survey outcome $y_{ij}$ for unit $i$ in cell $j$ follows a normal distribution with cell-specific mean $\theta_j$ and variance $\sigma_j^2$ values,
\begin{align}
\label{model}
(y_{ij} \mid \theta_j, \sigma_j)\sim N(\theta_j, \sigma_j^2),
\end{align}
the proposed MRP estimator for the population mean can be expressed in the weighted form,
\begin{align}
\label{mrp-g}
	\tilde{\theta}^{\,{\rm mrp}}=\sum_{j=1}^J\frac{N_j}{N}\tilde{\theta}_j,
\end{align}
where $\tilde{\theta}_j$ is the model-based estimate of $\theta_j$ in cell $j$. 

Under a Bayesian paradigm, given an exchangeable prior distribution, $(\theta_j \mid \mu, \sigma_\theta) \sim N(\mu, \sigma^2_{\theta})$, where the hyperparameter $\mu$ is assigned with a noninformative prior distribution, the posterior mean estimates $\tilde{\theta}_j$ are smoothed across cells, i.e., under partial pooling~\citep{gelman:hill-07}, resulting in the following MRP estimator given $(\sigma_j, \sigma_\theta)$,
\[
\tilde{\theta}^{\,{\rm mrp}}=\sum_{j=1}^J\frac{N_j}{N}\frac{\bar{y}_j+\delta_j\frac{\sum_{j=1}^J(\bar{y}_j/(1+\delta_j))}{\sum_{j=1}^J(1/(1+\delta_j))}}{1+\delta_j} \mbox{, where } \delta_j=\frac{\sigma_j^2}{n_j\sigma^2_{\theta}}.
\]
The ratio of sums $\frac{\sum_{j=1}^J(\bar{y}_j/(1+\delta_j))}{\sum_{j=1}^J(1/(1+\delta_j))}$ is a constant that does not depend on $j$ and can be approximated by $\sum_j\bar{y}_j*n_j/n=\bar{y}_s$. Hence, the approximated MRP estimator becomes
\begin{align}
\label{mrp}
\tilde{\theta}^{\,{\rm mrp}}\approx \sum_{j=1}^J\frac{N_j}{N}\frac{\bar{y}_j+\delta_j\bar{y}_s}{1+\delta_j} \mbox{, where } \delta_j=\frac{\sigma_j^2}{n_j\sigma^2_{\theta}},
\end{align}
as a combined estimator of $\bar{y}_{ps}$ and $\bar{y}_{s}$ with the shrinkage factor $\delta_j$, which is a function of the sample cell size $n_j$, within-cell variance $\sigma^2_j$ and between-cell variance $\sigma^2_{\theta}$. When $n_j \rightarrow 0$, $\delta_j \rightarrow \infty$, i.e., with small cell sizes, the estimate will be pooled toward the overall unweighted mean $\tilde{\theta}^{\,{\rm mrp}}\rightarrow \bar{y}_s$, as illustrated by the empirical work in~\cite{buttice_highton}; when $n_j \rightarrow \infty$, $\delta_j \rightarrow 0$, i.e., with large cell sizes, the estimate will be led toward the PS estimator with cell-wise direct estimates $\tilde{\theta}^{\,{\rm mrp}}\rightarrow \bar{y}_{ps}$.

Next, we examine the bias and variance by comparing UnW, PS, and MRP estimators. Assume that the sample respondents are a random sample of the population respondents: $E(\bar{y}_j)=\bar{Y}_{jR}$, similar to~\cite{Kalton88}, and we can calculate the bias

\[
\textrm{bias}(\bar{y}_{ps})= \sum_{j=1}^J\frac{N_j}{N}(1-\psi_j)(\bar{Y}_{jR}-\bar{Y}_{jM})\equiv \mathcal{B}
\]

\[
\textrm{bias}(\bar{y}_{s})=\sum_{j=1}^J\frac{N_j}{N}\frac{(\bar{Y}_{jR}-\bar{Y}_R)(\psi_{j}-\bar{\psi}) }{\bar{\psi}}+ \mathcal{B}\equiv \mathcal{A}+\mathcal{B},
\]
where $\bar{\psi}=\sum_{j=1}^J\frac{N_j}{N}\psi_{j}$ denotes the overall proportion of respondents in the population. The term $\mathcal{A}$ captures the variation between cell-wise response propensities, and the term $\mathcal{B}$ is a weighted average of cell-wise covariances between the response propensities and outcomes. The general conditions that the absolute values $|\textrm{bias}(\bar{y}_{ps})|<|\textrm{bias}(\bar{y}_{s})|$ are if: 1) $\mathcal{A}$ and $\mathcal{B}$ have the same sign or 2) the absolute values $|\mathcal{A}|>2|\mathcal{B}|$. The biases can be approximated in a stochastic model~\citep{Bethlehem02}
\[
\textrm{bias}(\bar{y}_{s})=\frac{\textrm{Cov}(\psi,y)}{\bar{\psi}}
\mbox{,  \hspace{0.5cm} }
\textrm{bias}(\bar{y}_{ps})=\sum_{j=1}^J\frac{N_j}{N}\frac{\textrm{Cov}_j(\psi,y)}{\psi_j},
\]
where $\textrm{Cov}(\psi,y)$ is the covariance between the response probabilities and the outcome values, and $\textrm{Cov}_j(\psi,y)$ is the covariance between the response probabilities and the outcome values within cell $j$. The expression $\textrm{bias}(\bar{y}_{s})$ is similar to the data defect index discussed in~\cite{meng2018}.

The bias of the MRP estimator, $\textrm{bias}(\tilde{\theta}^{\,{\rm mrp}})$, is given by
\begin{align*}
&\sum_{j=1}^J \frac{N_j}{N}\frac{1}{1+\delta_j}(1-\psi_j)(\bar{Y}_{jR}-\bar{Y}_{jM}) +\\
&\sum_{j=1}^J\frac{\delta_j}{1+\delta_j}\left(\sum_{j=1}^J\frac{N_j\psi_j}{N\bar{\psi}}\bar{Y}_{jR}-\psi_{j}\bar{Y}_{jR}-(1-\psi_{j})\bar{Y}_{jM}\right).
\end{align*}

If the dataset is an SRS, i.e., MCAR, $\psi_j\equiv\bar{\psi}$, then all three estimators are unbiased. When the cells are homogeneous with respect to either the response probability or the outcome variable, equivalently, based on the model~\eqref{model}, $\bar{Y}_{jR}=\bar{Y}_{jM}$, $\textrm{Cov}_j(\psi,y)=0$, and thus, $\mathcal{B}=0$, the PS estimator $\bar{y}_{ps}$ is unbiased. However, with the exchangeable prior distribution, the MRP estimator $\tilde{\theta}^{\,{\rm mrp}}$ will be biased, the second term of which is nonzero, even though $\textrm{bias}(\tilde{\theta}^{\,{\rm mrp}})$ will be smaller than $\textrm{bias}(\bar{y}_{s})$.

Conditional on the sample cell sizes $\vec{n}=(n_1,\dots, n_J)$, the variance estimates are
\begin{align*}
\textrm{var}(\bar{y}_{s}|\vec{n})&=\sum_{j=1}^J\frac{n_j}{n^2}s_j^2\\
\textrm{var}(\bar{y}_{ps}|\vec{n})&=\sum_{j=1}^J\frac{N_j^2}{N^2}(1-\frac{n_j}{N_j})\frac{s_j^2}{n_j},
\end{align*}
where $s_j^2$ is the element variance inside cell $j$. The conditional variance estimate for $\tilde{\theta}^{\,{\rm mrp}}$ given $\vec{n}$ and $\vec{\delta}=(\delta_1,\dots, \delta_J)$, $\textrm{var}(\tilde{\theta}^{\,{\rm mrp}}|\vec{n},\vec{\delta})$, is approximated by
\begin{align*}
\sum\frac{N^2_j}{N^2}\left((\frac{1}{1+\delta_j})^2\frac{s_j^2}{n_j}+
(\frac{\delta_j}{1+\delta_j})^2\frac{n_j}{n^2}s_j^2+\frac{2\delta_j}{(1+\delta_j)^2}\frac{s_j^2}{n}\right).
\end{align*}
With small $n_j$'s, the variance $\textrm{var}(\bar{y}_{ps}|\vec{n})$ could be large, and $\textrm{var}(\tilde{\theta}^{\,{\rm mrp}}|\vec{n},\vec{\delta})$ is reduced due to the shrinkage effect. Under a Bayesian paradigm, we can obtain the MRP variance estimate, $\textrm{var}(\tilde{\theta}^{\,{\rm mrp}}|\vec{n})$, after marginalizing $\vec{\delta}$ in the posterior computation.

Under MRP, the mean estimates for small cells will be pooled toward the overall mean when the between-cell variation is not well captured. Normal prior distributions on $\theta_j$ with exchangeability can result in over-shrinking and potentially biased estimates. The over-shrinking problem can be avoided by including more auxiliary information or alternative prior specifications such as the global-local shrinkage priors~\citep{priorSAE-Tang18}. 

Here, we focus on the incorporation of predictive cell-wise covariates to introduce a conditionally exchangeable model for $\theta_j$ given the auxiliary variables $X_j$, for example,
$
\theta_j = X_j\beta + \gamma_j,
$
where $\gamma_j \indep X_j \mbox{, } (\gamma_j \mid \sigma_\theta) \sim N(0, \sigma^2_\theta)$. We use cell indices $j$ in the notation of auxiliary variables $X_j$, rather than unit indices $i$, because the poststratification cells are the resulting cross-tabulation of $X_i$. It is appropriate to assume a conditional exchangeable model after incorporating sufficient relevant information in $X_j$ in the mean structure of $\theta_j$ that the cells can be considered as randomly assigned. With cell-wise covariates $X_j$, the MRP estimator, which is commonly used in practice, becomes
\begin{align}
\label{mrp-x}
\tilde{\theta}^{\,{\rm mrp}}\approx \sum_{j=1}^J\frac{N_j}{N}\frac{\bar{y}_j+\delta_jX_j\beta }{1+\delta_j} \mbox{, where } \delta_j=\frac{\sigma_j^2}{n_j\sigma^2_{\theta}},
\end{align}
as a combined estimator of $\bar{y}_{ps}$ and $\sum_{j=1}^J\frac{N_j}{N}X_j\beta$. The MRP estimator is a population average of the composite estimators popular in small area estimation~\citep{Ghosh:sae2020}.

\cite{Lahiri:AOS07} have examined the design-consistency property of a hierarchical Bayes estimator of a finite population stratum mean when the sample size is large. Specifically, they prove that with an exponential family distribution for the outcome and a normal prior distribution for $\theta_j$, the posterior mean estimator $\tilde{\theta}_j \rightarrow \bar{y}_j$ as $n_j \rightarrow \infty $ and $n_j/N_j \rightarrow f_j$ for some $0<f_j<1$, and the corrected estimator $\tilde{\theta}_j - (\bar{y}_j - \bar{y}_{j,ds})$ is design-consistent, where $\bar{y}_{j,ds}$ is any design-consistent estimator of $\bar{Y}_j$ (Theorem 3.1, ~\cite{Lahiri:AOS07}). Extending to MRP, as a combined estimator of $\bar{y}_{ps}$ and $\bar{y}_{s}$, we have the corollary below.

\begin{corollary} Assume the following regularity conditions.\\
(R.1) The survey outcome $y_{ij}$ follows an exponential family distribution with cell-specific mean $\theta_j$ and variance parameters $\sigma^2_j$.\\
(R.2) The prior distribution of $\theta_j$ is $N(\mu_j, \sigma^2_\theta)$.\\
(R.3) The cell-wise proportion $n_j/N_j \rightarrow f_j$ for some $0<f_j<1$.\\
(R.4) The poststratification cell structure fully accounts for the design information and nonresponse.\\
Then the MRP estimator
\[
\tilde{\theta}^{\,{\rm mrp}}=\sum_{j=1}^J\frac{N_j}{N}\tilde{\theta}_j \rightarrow \sum_{j=1}^J \frac{N_j}{N}\bar{y}_j,
\] 
as $n_j \rightarrow \infty $, is design-consistent.
\end{corollary}

Design consistency is desirable in the randomization approach to finite population sampling. One of the critical regularity conditions is the availability of auxiliary information that constructs the poststratification cells and fully accounts for the design information and nonresponse mechanism. Moreover, MRP improves estimation efficiency. With data-driven shrinkage modeling to strike a tradeoff between bias and variance, especially with a flexible model structure and suitable prior distributions, we expect that $\tilde{\theta}^{\,{\rm mrp}}$ can yield the smallest root mean squared error (RMSE) and robust inferences.   

\subsection{Robust inferences}
\label{robust}

Below we discuss the connections of MRP with alternative methods and highlight the improvement via the selection bias adjustment and hierarchical modeling to be robust against misspecification. The crucial guidelines on the use of auxiliary information under MRP include 1) poststratification with predictive auxiliary information and 2) flexible outcome modeling strategies.

{\bf Poststratification with predictive auxiliary information:} \cite{wang:gelman14} have applied MRP to the 2012 U.S. presidential election forecasting based on a non-representative poll of 350,000 Xbox users and calibration with the exit poll data. Their findings show that MRP estimates are in line with the forecasts from leading poll analysts. The success of MRP mostly comes from the adjustment for differential nonresponse and turnouts to which vote swings are mostly attributed with highly predictive auxiliary information for voting behavior~\citep{gelman-swing16}.

Incorporating auxiliary variables as covariates in response propensity modeling, inverse propensity score weighting (IPW) estimators use the inverse of predicted propensities as weights but are sensitive to misspecification, especially with a wide range of weight values that can result in highly variable estimates (e.g., ~\cite{Tan07}). Flexible predictive models are fit to construct stabilized pseudo-weights for nonprobability samples, such as Bayesian additive regression trees~\citep{Bart-Elliott20} and kernel weighting approaches~\citep{kernal-Li20}. In contrast to IPW, MRP constructs poststratification cells based on the discretized auxiliary information and groups individuals under multilevel modeling to reduce the variability. 

Either when 1) the inclusion probabilities of the individuals are the same within cells, or 2) the included individuals are similar to those excluded within cells, i.e., conditional MAR within cells, the poststratification and MRP estimators will be unbiased. \cite{holt:smith:79} show that poststratification based on cells that are homogeneous with respect to the target variable reduces both variance and bias. Poststratification based on cells that are homogeneous with respect to the response probabilities reduces the bias but not necessarily the variance~\citep{little-nonresponse-ISR86}. Consistent with~\cite{little-weighting-SM2005}, we recommend incorporating auxiliary variables that are predictive of either the survey outcomes (primarily) or response propensities (secondly) in the construction of poststratification cells. With a large number of auxiliary variables, the number of poststratification cells $J$ could be large, for example, $J=2\times 4\times4\times 5 \times 50=8000$, in the adjustment of sex (2 levels), race/ethnicity (4 levels), education (4 levels), age (5 levels) and state (50 levels). The resulting sample poststratification cells could be sparse and even empty, as the computational bottleneck for MRP. \cite{little93} recommends collapsing small cells to reduce the variance, with a payoff of increased bias. The Bayesian paradigm of MRP allows data to determine the pooling effects and stabilize inferences with a balance between bias and variance.

{\bf Flexible outcome modeling strategies:} DR estimators improve IPW estimators by combining with a prediction model for the survey outcome~\citep{bang:robins05}. \cite{kang:schafer07} show that the regression prediction estimator outperforms DR estimators with a predictive model. Moreover, most DR estimators do not apply to subgroup estimation. With a linear regression model, the general regression (GREG) estimator improves estimation accuracy in combination with poststratification adjustments~\citep{greg92}. MRP replaces the linear regression in GREG with a multilevel model with smoothing effects across poststratification cells. 

The flexible models under MRP improve small area estimation and facilitate finite population inferences, with examples including Bayesian hierarchical models with high-order interactions and global-local shrinkage prior specifications~\citep{Ghitza:gelman:13,prior-si2018}, Gaussian process (GP) regression models~\citep{BNFP:SI15}, stacked regression~\citep{stackMRP20}, and machine learning algorithms~\citep{automrp:22}. 

\subsection{Systematic data integration}
\label{inte-mrp}

Despite numerous successful applications, MRP is faced with challenges in practical settings. \cite{Valliant19-jssam} and \cite{Lau2018} show that the MRP estimates of population means are biased and inferior to alternatives. We replicate their studies and find that the covariates in the model of \cite{Valliant19-jssam} are not strongly related to the outcome, resulting in underestimated between-cell variation and cell-wise estimates over-shrunk toward the overall mean, consistent with our discussion in Section~\ref{method-ps}. MRP will fail when the outcome model is not correctly specified, for example, with non-predictive covariates or strong prior distributions over-shrinking the between-cell variance parameters toward zero. We propose an integrated framework to improve the MRP estimator in settings such as those of \cite{Valliant19-jssam}, which will also be illustrated in the simulation studies of Section~\ref{simulation}.

Current approaches for inferences with nonprobability survey samples assume that the relevant auxiliary information is available from a reference probability survey sample~\citep{elliott-valliant17,wu2018,Yang:Kim20,Kim:Tam20}. In practice, the reference probability sample may be too small to fully recover the required population characteristics. Another issue is empty cells that can occur in either nonprobability samples or the reference sample, ignoring which may lead to biased population estimates. Most existing MRP applications treat the population cell counts $N_j$'s from large-scale external data (e.g., ACS and CPS) as fixed and ignore the sampling uncertainty. We would like to generate synthetic population information if it is unknown or estimated from a small reference sample, as a data integration framework on the use of auxiliary information in MRP. The general MRP estimator for the population mean would be
\begin{align}
\label{g-mrp}
	\tilde{\theta}^{\,{\rm mrp}}=\sum_{j=1}^J\frac{\hat{N}_j}{\sum \hat{N}_j}\tilde{\theta}_j,
\end{align}
where both $\hat{N}_j$ and $\tilde{\theta}_j$ are estimates.

A variety of nonparametric and parametric approaches are available to estimate $\hat{N}_j$. With survey weights from the probability sample, weighted finite population Bayesian bootstrap (WFPBB) approaches have been developed to obtain the estimated population distribution of $X_j$~\citep{zangeneh:little12,fpbb14,Zhou:Biometrics16,Makela-sm18}. Often the relevant auxiliary information is partially collected and requires modeling to generate the synthetic population distribution~\citep{reilly:gelman:katz01,unknownx:mrp:Phillips15,emrp21}. \cite{BNFP:SI15} have accounted for survey weights in probability samples in a multinomial model to yield smoothed estimates of $\hat{N}_j$'s with sparse cells. \cite{mrsp:leemann17} combine MRP with synthetic populations when only population margins of the auxiliary variables are known and apply bootstrap algorithms to account for the estimation uncertainty. Their approach is one type of model-based raking adjustments where the auxiliary variables have independent effects on the inclusion mechanism~\citep{rake:little91,BayesRake18}, but different from the conventional raking based on the iterative proportional fitting algorithm that focuses on minimizing specified distance functions and ignores the uncertainty.  

The availability and selection of poststratification cells influence the performance of MRP. When only observed cells in a nonprobability sample are used, empty cells are not handled, despite weighting and poststratification methods relying on available cells. We consider two MRP methods: MRP-P, which predicts outcomes for all population cells using known population cell sizes, and MRP-R, which uses available cells in the reference probability sample with unknown population cell sizes to account for estimation uncertainty. MRP-P is commonly used in practice.

The inclusion probabilities can be predicted by solicited auxiliary variables for all population units under the quasi-randomization assumption of inclusion into the nonprobability sample. We recommend incorporating a flexible function form of estimated inclusion probabilities as a predictor to borrow information across cells, in addition to carefully selecting predictive auxiliary variables and their high-order interaction terms for the outcome modeling. This will improve the outcome model fitting and facilitate robustness under the assumption of homogeneous poststratification cells. To this end, we propose an integrated MRP approach, MRP-INT, which adds estimated inclusion probabilities as predictors in the outcome model and predicts all population cells using known population cell sizes. MRP-INT combines quasi-randomization and superpopulation modeling, and involves constructing three main models under the data integration framework.

{\bf 1. Cell sizes}: assume that the respondents within poststratification cell $j$ are independently and quasi-randomly sampled from the population cell cases,
\begin{align}
\label{nj}
(n_1,\dots,n_J \mid \vec{\psi}, \vec{N}) \sim \textrm{Multi}\left((cN_1\psi_1,\dots, cN_J\psi_J), n\right),
\end{align}
where $\vec{N}=(N_1,\dots, N_J)$, $\vec{\psi}=(\psi_1, \dots, \psi_J)$, and $c=1/\sum_{j=1}^JN_j\psi_j$, a normalizing constant. The multinomial ($\textrm{Multi}$) distribution can be approximated by a series of $J$ Poisson distributions when $J$ is sufficiently large or the inclusion probability $\psi_j$ is sufficiently small.

{\bf 2. Inclusion probabilities}: we can assume that they are concentrated around the average inclusion probability of the sample of interest, for example, a Beta distribution with a mean of $n/N$. Alternatively, with covariates $X^{\psi}_j$ that affect inclusion propensities and a link function $g(\cdot)$ (e.g., {\em logit}), we model the inclusion probabilities used in~\eqref{nj} and smooth the estimates across cells,
\begin{align}
\label{pj}
g(\psi_j)=X^{\psi}_j\alpha,
\end{align}
where $\alpha$ denotes coefficients, and $X^{\psi}_j$ is the list of variables $X_j$ that predict the inclusion probability. 

{\bf 3. Outcome}: within each poststratification cell, the units are included with the same probability and independently distributed. We assume that the outcome depends on the cell inclusion probabilities and follows a normal distribution with cell-specific mean and variance values,
\begin{align}
\label{y-j}
(y_{ij} \mid \psi_j, X^y_j, \beta, \sigma_j) \sim N\left(f(\psi_j) + X^{y}_j\beta, \sigma_j^2\right).
\end{align}
Here $f(\psi_j)$ is a function of $\psi_j$, and $X^{y}_j$ denotes the cell-wise covariates as a list of auxiliary variables $X_j$ that predict the outcome variable, may include both main effects and high-order interaction terms, and overlap with $X^{\psi}_j$. For notational simplicity, we ignore individual-level covariates, which are predictive of the outcome and can even be continuous, although not used in the poststratification. With rich information from $X^{y}_j$ and correct specification of the mean structure $X^{y}_j\beta$, the role of the function $f(\psi_j)$ could be minimal. We recommend including the flexible specification of $f(\psi_j)$ that offers protection against model misspecification, similar to the doubly robust penalized spline of propensity prediction methods~\citep{little:an04,breidt05,zhang:little09}. 

Flexible prior distributions can be introduced on $f(\psi_j)$, such as normal distributions over discrete categories based on the unique $\psi_j$ values, penalized spline functions and nonparametric Bayesian distributions, to account for the dependency structure and smoothing effects across cells. \cite{BNFP:SI15} construct poststratification cells based on unique $\psi_j$ values and induce a GP prior distribution, i.e., $f(\psi_j) \sim GP(\mu(\psi_j), \Sigma(\vec{\psi}))$, where the mean $\mu(\psi_j)$ is a linear function of the logarithm of the inclusion probability, and the covariance matrix $\Sigma(\vec{\psi})$ depends on the differences between $\psi_j$'s, e.g., $\textrm{Cov}(\psi_j,\psi_{j'})=(1-g)\tau^2exp(-\frac{(\psi_j-\psi_{j'})^2}{l^2}) +g\tau^2I_{j={j'}}$, with weakly informative or noninformative prior specifications for hyperparameters $(g, \tau, l)$. The smoothing function $f(\psi_j)$ can be pre-specified or approximated with basis expansion functions if the number of cells is large. 


The computation for models specified in~\eqref{nj},~\eqref{pj}, and~\eqref{y-j} can be implemented in a fully Bayesian procedure or multiple sequential steps. Stan makes MRP computation generally accessible and can perform Markov chain Monte Carlo (MCMC) computation. The implementation of MRP is straightforward with the publicly available R packages such as {\em Rstan}~\citep{rstan:2020}, and {\em rstanarm}~\citep{rstanarm}. For example, in {\em rstanarm}, the function {\em stan\_glmer} fits a multilevel model and the function {\em posterior\_predict} imputes the outcome for all nonsampled units in the population. The imputation step draws posterior predictive samples and generates multiple synthetic populations. We can also apply poststratification to the posterior samples. Subgroup estimates are obtained by extracting imputed values for the corresponding domains. Even with scalable algorithms in Stan, the MCMC computation of the prediction modeling is more time-consuming than model-free weighting approaches.

Table~\ref{mrp-sum} summarizes different MRP approaches, depending on the outcome model specification, the selection of cells for poststratification, and whether the population cell sizes $N_j$ are known or estimated. 

\begin{table}
\caption{Comparison of different multilevel regression and poststratification (MRP) approaches.}
\label{mrp-sum}
\center
\begin{tabular}{llll}
\hline
& Mean& Poststratification& Population \\
& & cells& cell size \\
\hline
MRP-P & $X^{y}_j\beta$ & population & known\\
MRP-R & $X^{y}_j\beta$ & reference sample& estimated\\
MRP-INT & $f(\psi_j) + X^{y}_j\beta$ & population& known\\
\hline
\end{tabular}
\end{table}

\section{Simulation studies}
\label{simulation}

In this section, we compare different MRP approaches with the IPW, GREG, DR, and raking adjustments. To demonstrate the effectiveness of the proposed data integration framework, we use simulation setups similar to those in \cite{Valliant19-jssam} and show how MRP fails and our proposed framework offers improvement. More details about the simulation design and results can be found in the Supplementary Material~\citep{mrp:supp:ys22}.

We simulate a population of size $N=50,000$ based on the data {\em mibrfss} in the R package PracTools~\citep{Valliant:practoolsR}. This dataset was collected through telephone household interviews in the Michigan Behavioral Risk Factor Surveillance Survey. We randomly sample the {\em mibrfss} data of 2,845 cases with replacement to generate the population. Among the population, 65.5\% have access to the internet at home. 

We draw nonprobability samples from the subset of the population with internet access, while reference probability samples are SRSs drawn from the whole population. We use demographic variables: age in years (1 = 18--24; 2 = 25--34; 3 = 35--44; 4 = 45--54; 5 = 55--64; 6 = 65+), race (1 = White; 2 = African American; 3 = Other), edu: education level (1 = Did not graduate high school; 2 = Graduated high school; 3 = Attended college or technical school; 4 = Graduated from college or technical school), and inc: income (1 = Less than \$15,000; 2 = \$15,000 to less than \$25,000; 3 = \$25,000 to less than \$35,000; 4 = \$35,000 to less than \$50,000; 5 = \$50,000 or more). The internet subgroup has higher proportion of younger individuals than the population. The nonprobability samples are selected from the internet subpopulation with age categories as strata. The selection rates across six strata are set as $(0.12, 0.31, 0.19, 0.20, 0.13, 0.05)$, with younger individuals more likely to be sampled than the eld. 

The outcome variables in \cite{Valliant19-jssam} are collected from the {\em mibrfss} data, and we find that the model fitting performances are not satisfactory in the approaches that use prediction models. Hence, we simulate an outcome variable $Y$ from different pre-specified models and implement two simulation scenarios: one with a correctly specified model and the other with an incorrect specification, for all the evaluated methods.

In the MRP approach, we use the four covariates to construct poststratification cells, main effects of the four variables in the outcome model, and the default weakly informative prior choices for regression models in the R package {\em rstanarm}. The WFPBB approach generates 100 synthetic populations with the same size $N=50,000$, i.e., 100 posterior samples of $\hat{N}_j$'s for poststratification under MRP-R. We include main effects of four covariates in the model~\eqref{pj} and combine it with model~\eqref{nj} to estimate the inclusion probabilities for MRP-INT. In the outcome model~\eqref{y-j} under MRP-INT, in addition to including the estimated inclusion probabilities $\psi_j$'s after a logit transformation as a numeric predictor, we create discrete categories via one-to-one mapping of the unique $\psi_j$ values rounded with one digit, similar to \cite{BNFP:SI15} assuming that units within the same inclusion probabilities are assigned to the same group, and introduce a varying intercept across groups.

The IPW method fits a weighted logistic regression model with four main effects to the concatenated nonprobability and reference samples and uses the inverse of the predicted probabilities of belonging to nonprobability samples as weights. Note that this probability, in the nonprobability sample vs. in the probability sample, is different from our defined inclusion probability $\psi$, in the nonprobability sample vs. not in the nonprobability sample for the population units. The prediction models in GREG, DR, and raking estimators include main effects of the four variables. The GREG fits a linear prediction model given population margins, and the DR estimator combines GREG predictions with IPW. Raking estimators account for known population margins with the iterative proportional fitting algorithm. 

The quantities of interest are the overall population mean and subgroup means across six age groups. The six groups have different selection probabilities and population sizes ranging from $2,918$ to $10,860$. We use the posterior samples generated by MRP methods to obtain the mean, standard error (SE) estimates, and credible intervals treated as confidence intervals (CIs). We estimate the variance of IPW, GREG, DR, and raking estimators with the jackknife replication approach. We repeatedly draw the nonprobability and reference probability samples from the population and evaluate the frequentist properties by calculating the relative bias, RMSE, average SE estimates, and nominal coverage rates of 95\% confidence intervals. The relative bias is defined as the ratio of estimated bias to the true value, $\frac{\sum_s\hat{\theta}_s/S - \theta}{\theta}$, where $\hat{\theta}_s$ is the estimate of the true value $\theta$ from sample $s$, for $s=1,\dots,S$.

We conduct 200 repeated studies under three MRP approaches and 1000 times under alternative methods, including IPW, GREG, DR, and raking. The MCMC computation and synthetic population generation are computationally intensive. For each replication, we draw a stratified sample of size $n=1000$ from the internet subpopulation as the nonprobability study of interest and an SRS sample of size $n=1000$ from the entire population as the reference probability sample.

\subsection{Correct model specification}
\label{sim-case1}

The population data generation model for the outcome $Y$ is a normal distribution with a standard deviation of 1, and the mean structure depends on the main effects (age, race, edu, inc), for which we randomly sample integer values between -5 and 5 with replacement and set them as coefficients. In this setup, the outcome models under MRP-P, MRP-R, GREG, and DR are correctly specified, and the model to construct weights in IPW and DR, as well as raking, includes the correct terms.

\begin{figure}
\begin{tabular}{c}
\includegraphics[width=0.95\textwidth]{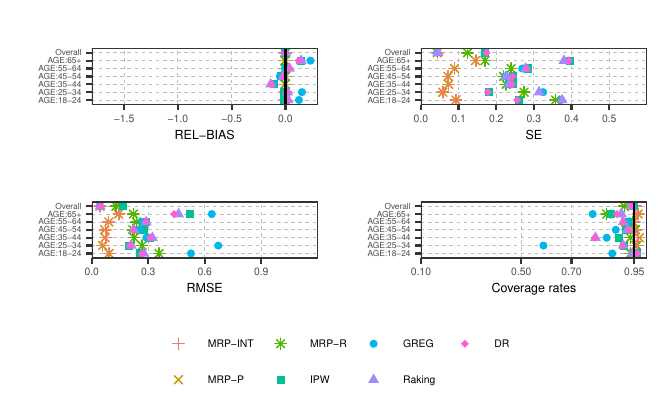}
\end{tabular}
\caption{\em \small Outputs of overall and subgroup mean estimates when the outcome models and adjustment factors of all methods are correctly specified. RMSE: root mean squared error, DR: doubly robust estimator, IPW: inverse propensity weighting estimator, GREG: generalized regression estimator, MRP: multilevel regression and poststratification, MRP-INT: an integrated MRP by adding estimated inclusion probabilities as predictors in the outcome models and predicting all population cells with known population cell sizes, MRP-P: MRP that uses all population cells with known population cell sizes, and MRP-R: MRP that uses available cells in the reference probability sample with unknown population cell sizes. }
\label{sim-y}
\end{figure}

The comparison results are presented in Figure~\ref{sim-y}. All methods are unbiased or have negligible biases. MRP-P and MRP-INT exhibit the best performances, particularly for subgroup estimates. The estimation model under MRP-P is correctly specified, and all population cell information is fully utilized. The outcome model under MRP-INT covers the correct model as a special case, and the effect of inclusion propensities is shrunk toward zero without loss of efficiency. Using the reference sample to estimate the population cell sizes can correct the sample selection bias but result in considerable variability given by the MRP-R estimates. The WFPBB algorithm accounts for the uncertainty in the population cell size estimation and can yield noisy estimates, especially when the reference sample is small. The IPW yields reasonable coverage rates but has larger SE and RMSE values than MRP-P and MRP-INT estimates. Raking also has large variability but slight under-coverage. Model-based approaches tend to have lower SEs than the IPW and raking, but not for all subgroup estimates. The GREG estimates of the overall mean have good properties, but the subgroup mean estimates have large RMSE values and low coverage rates. Moreover, the linear model prediction results in negative weight values, which can be potentially resolved by imposing bounds on the minimization problem whose solution produces the weights, but we ignore the bounds here in the comparison. DR generates similar SEs and RMSEs to those of IPW and raking. When neither weighting nor prediction approaches work well, DR estimates do not present improvements. Overall, with correct outcome models, MRP-P and MRP-INT outperform alternative approaches with efficiency gains on subgroup mean estimates.

\begin{figure}
\begin{tabular}{c}
\includegraphics[width=0.95\textwidth]{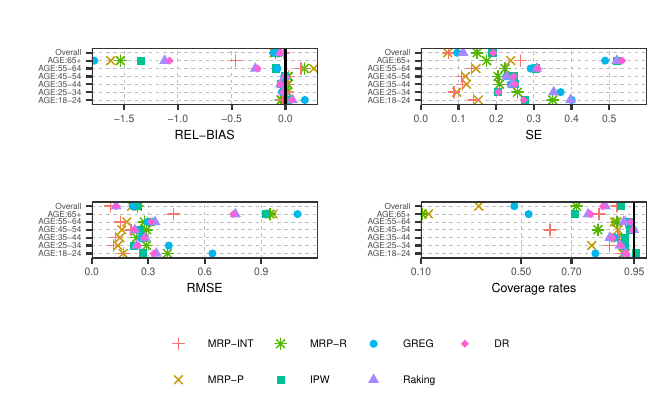}
\end{tabular}
\caption{\em \small Outputs of overall and subgroup mean estimates when the outcome models and adjustment factors of all methods are incorrectly specified. RMSE: root mean squared error, DR: doubly robust estimator, IPW: inverse propensity weighting estimator, GREG: generalized regression estimator, MRP: multilevel regression and poststratification, MRP-INT: an integrated MRP by adding estimated inclusion probabilities as predictors in the outcome models and predicting all population cells with known population cell sizes, MRP-P: MRP that uses all population cells with known population cell sizes, and MRP-R: MRP that uses available cells in the reference probability sample with unknown population cell sizes. }
\label{sim-y-main}
\end{figure}

\subsection{Incorrect model specification}
\label{sim-case2}
 
We extend the data generation model for $Y$ above in Section~\ref{sim-case1} by adding two-way interactions (age * edu, race * inc), with coefficients randomly sampled from integers between -5 and 5. Consequently, the outcome models with only main effects in all methods are misspecified, and IPW, raking, and DR do not account for the interactions in the weighting adjustment. 

Figure~\ref{sim-y-main} shows that MRP-P fails to yield reasonable coverage rates for the overall mean and most subgroups due to the model misspecification. MRP-P has a large bias, high variability, and low coverage for the subgroup estimate of those aged 65 years and older. MRP-INT improves MRP-P by providing additional prediction power through inclusion probabilities, resulting in smaller bias and SE values and larger coverage rates. However, the coverage rate under MRP-INT for one subgroup estimate (Age 45-54) is 0.60, indicating invalid inference.  Although MRP-INT improves subgroup inferences, it is not effective for all estimates. This is because the proposed model adding estimated inclusion probabilities is still different from the true model that involves two interaction terms. Nevertheless, the formulation provides protection against model misspecification. 

MRP-R has larger SE and RMSE than MRP-P, as it accounts for variability in estimating population cell sizes. However, its coverage rates are below the nominal levels for all estimates. The bias of MRP-R is compatible with MRP-P because the outcome model only has main effects. The variability of MRP-R can be substantial when the outcome model becomes complex, such as including numerous high-order interactions, where the reference sample fails to recover the population structure. WFPBB is able to recover the marginal distributions in the population but not necessarily the joint distribution, leading to substantial variation in the poststratification adjustment. The alternative methods, including the IPW, GREG, DR, and raking, also exhibit poor performances. All methods perform similarly for the overall mean estimate but with low coverage rates. GREG has high variability and low coverage for the group estimate of 65 years and older. Again GREG results in negative weights.

The values for Figures~\ref{sim-y} and ~\ref{sim-y-main} are presented in the tables of the Supplementary Material~\citep{mrp:supp:ys22}.

\section{Application}
\label{application}

We apply MRP to make inferences from our motivating nonprobability survey, the ABCD study. The ABCD study is a prospective cohort study and has collected social, health, imaging, and genetics data from 11,875 children aged 9-10 from 21 U.S. research sites between 2016 and 2018. The ABCD study's sampling and recruitment process aims to yield an overall sample that closely matches national sociodemographics of U.S. children aged 9--10~\citep{ABCDdesign}. However, the 21 research sites were not randomly chosen but rather selected based on convenience such as neuroimaging resource allocation and accessibility, resulting in potential selection bias. Moreover, child enrollment is conditional on the school and parental consent, and low consent rates can further exacerbate nonresponse bias.

\begin{table}[h!]
\center
\caption{Sociodemographic distribution comparison between the Adolescent Brain Cognitive Development (ABCD) baseline cohort (March 2019, $n=11,875$) and the American Community Survey (ACS, 2011--2015, $N=8,211,605$, adjusted by the ACS weights). Values are in percentage.}
\label{abcd-acs}
\footnotesize
\begin{tabular}{lll}\\\toprule  
&ABCD&ACS\\
\hline
\rowgroup{\em Age}&&\\
9&56.6&49.6\\
10&43.4&50.4\\
\rowgroup{\em Sex}&&\\
Male	&52.1&51.2\\
Female&47.9&48.8\\
\rowgroup{\em Race/Ethnicity}&&\\		
White&52.1&52.4\\
Black&15.0&13.4\\
Hispanic&20.3&24.0\\
Asian&2.1&5.9\\
Other&10.5&4.2\\
\rowgroup{\em Family Type}&&\\				
Married couple&73.1&66.1\\
Other&26.9&33.9\\
\rowgroup{\em Household Size}&&\\			
2-3 Persons&17.5&18.5\\
4 Persons&33.7&33.5\\
5 Persons&25.2&25.4\\
6 Persons&14.3&12.5\\
7 or more Persons&9.3&10.1\\
\rowgroup{\em Household Income}&&\\			
$ < 25$K&16.0&21.5\\
\$25K--\$49K&15.2&21.7\\
\$50K--\$74K&13.7&17.0\\
\$75K--\$99K&14.2&12.5\\
\$100K--\$199K&29.9&20.5\\
\$200K+&11.0&6.8\\
\rowgroup{\em Family and Labor Force (LF) Status}&&\\			
Married, both in LF&49.9&40.8\\
Married, 0/1 in LF&23.2&25.6\\
Single parent, in LF&21.2&26.5\\
Single parent, not in LF&5.7&7.0\\
\bottomrule
\end{tabular}
\end{table}

Table~\ref{abcd-acs} compares the sociodemographic compositions between the ABCD study sample and the ACS 2011-2015 data. We find that the ABCD study oversamples 9-year old, males, high-income families, and certain race/ethnicity groups, with slightly more representation of children from families with married couples and being employed. By design, while the race/ethnicity composition for major classes (e.g., White and Black) matches the ACS targets fairly closely, children of Asian ancestry are underrepresented, and children with self-reported other race /ethnicity (AIAN, NHPI, Multiple) are overrepresented relative to the U.S. population of 9- and 10-year olds. The sociodemographic discrepancies from the U.S. population can potentially introduce selection bias analyzing data~\citep{abcd:bias19}.

To address this issue, \cite{ABCD:Heeringa-baselinewght} have developed weights that adjust for the sociodemographic differences between the ABCD sample and the ACS population control information. The weights are constructed by predicting pseudo-probabilities of sample inclusion and performing raking adjustments. Specifically, a multiple logistic regression model was fit to the concatenated ACS and ABCD data with covariates: age, sex, race/ethnicity, family income, family type, household size, parents' labor force status, and census region. The model predicted the pseudo-probabilities of inclusion in the ABCD sample (vs. in the ACS sample), the inverse of which were treated as the initial weights, in the same manner as the IPW in Section~\ref{simulation}. The initial weights were trimmed at the 2\% and 98\% quantiles of the distribution and calibrated with raking adjustments of age, sex, and race/ethnicity, the marginal distributions of which were matched to the ACS. However, the final weights are still widely spread and can result in unstable estimates.

In contrast to IPW, we apply three MRP approaches to adjust for the sample discrepancies in estimating average cognition test scores for the overall U.S. population of 9- and10-year olds and diverse sociodemographic groups of interest. The score is a total composite score of cognition based on various measures~\citep{ABCD:Heeringa-baselinewght}. Following literature studies~\citep{cognitive-Bradley02}, we use seven auxiliary variables in Table~\ref{abcd-acs} that are predictive of the cognitive outcome to construct poststratification cells. The cross-tabulation results in 4800 ($=2*2*5*2*5*6*4$) cells, 1517 of which are available in the ABCD data, and 3128 cells are available in the ACS data. There are 3 cells in the ABCD that are not available in the ACS. In the ABCD data, 962 cells have fewer than or equal to 5 units, and 333 cells only include 1 unit.

We can predict the outcome values for the 3128 cells in the ACS with the population counts. For MRP-P, we integrate these two datasets by fitting a model to the ABCD and then predicting the potential outcomes of the ACS dataset, where we treat ACS as the population data, account for the ACS weights in the estimates, but ignore the sampling error. For MRP-R, we treat ACS as a reference probability sample with survey weights and generate synthetic populations accounting for the sampling error. Since ACS is large, we expect that the extra uncertainty in MRP-R would be small compared to MRP-P.

The outcome model under MRP-P and MRP-R includes the main effects of seven auxiliary variables.
\begin{align*}
y_{ij} =& \beta_0 + \beta_{age}I(age_{j[i]}) + \beta_{fem}fem_{j[i]} + \beta_{mar}mar_{j[i]} +\\
& \beta^{race}_{j[i]}+\beta^{inc}_{j[i]} + \beta^{hhsize}_{j[i]} + \beta^{lf}_{j[i]} + \epsilon_i.
\end{align*}
for $i=1,\dots, n (=11,875)$, where $j[i]$ is the cell index that unit $i$ belongs to. Here we use dummy indicators for age (9, 10), sex (fem: female, male), and family type (mar: married, other), and multiple terms indicating levels for race/ethnicity (race), family income (inc), household size (hhsize), and family labor force status (lf), which are assigned with default weakly informative prior distributions in the R package {\em rstanarm} (Details are given in the Supplementary Material~\citep{mrp:supp:ys22}). 

In MRP-INT, we use weighted ACS to obtain the population cell sizes and include the estimated inclusion probabilities as predictors. We include main effects of seven variables above in the model~\eqref{pj} in combination with model~\eqref{nj} to estimate the inclusion probabilities of population units into ABCD. The MRP-INT outcome model~\eqref{y-j} includes the seven main effects, the estimated inclusion probabilities $\psi_j$'s after a logit transformation as a numeric predictor, and random effects of discrete categories via one-to-one mapping of the unique $\psi_j$ values rounded with one digit, similar to the setup in Section~\ref{simulation}.

We perform the computation in Stan with four MCMC chains of 5000 iterations each. The diagnostic measures $\hat{R}$ have values around 1 and indicate convergence. Under MRP-P and MRP-R, the covariates age, sex, race/ethnicity, family income, household size and labor force status are strong predictors of the test score, while the marital status is not predictive ($\hat{\beta}_{married}=-1.18$, 95\% CI $(-5.92, 2.85)$). The 10-year-olds tend to have a higher score than the 9-year-olds ($\hat{\beta}_{age}=4.59$, 95\% CI $(4.31, 4.86)$), and the girls have higher scores than boys ($\hat{\beta}_{fem}=0.59$, 95\% CI $(0.30, 0.87)$). The posterior median estimates of the standard deviation parameters are larger than 0, $\hat{\sigma}_{race} =10.54$ (95\% CI: $3.53, 54.54$), $\hat{\sigma}_{inc} =0.89$ (95\% CI: $0.21, 7.98$), 
$\hat{\sigma}_{hhsize} =13.54$ (95\% CI: $3.81, 84.69$), and $\hat{\sigma}_{lf} =0.91$ (95\% CI: $0.06, 39.56$), showing the covariates are predictive for the cognition outcome. Since these covariates have different distributions between ABCD and ACS, we expect that the adjustment to the ACS will generate different results from the unadjusted ABCD sample analysis. The MRP-INT outcome model presents the coefficient estimate of the inclusion probability $\hat{\beta}_{\psi} = 0.45$ with 95\% CI $(-1.11, 1.92)$ and the standard deviation of the group varying intercepts $\hat{\sigma}_{\psi} =0.07$ (95\% CI: $0, 0.26$). The effect of inclusion probabilities is not substantial.

\begin{figure}[ht]
\begin{tabular}{c}
\includegraphics[width=0.95\textwidth]{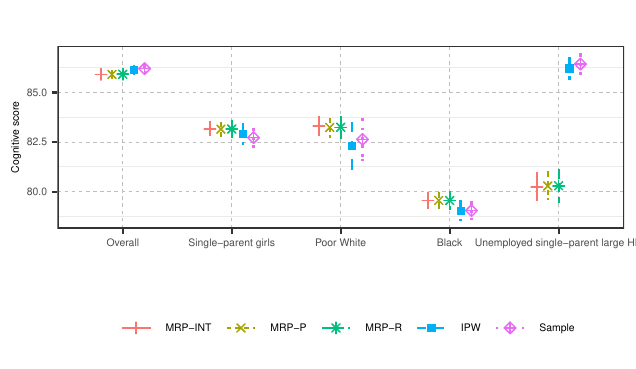}
\end{tabular}
\caption{\em \small Finite population inferences of average cognition test scores by groups, with seven auxiliary variables. The error bars are 95\% confidence intervals. HH: household, IPW: inverse propensity weighting estimator, MRP: multilevel regression and poststratification, MRP-INT: an integrated MRP by adding estimated inclusion probabilities as predictors in the outcome models and predicting all population cells with known population cell sizes, MRP-P: MRP that uses all population cells with known population cell sizes, and MRP-R: MRP that uses available cells in the reference probability sample with unknown population cell sizes. }
\label{abcd-fig}
\end{figure}

We also apply the pseudo-weights of the ABCD baseline survey and compute the weighted estimates, where the eligible sample size is $11,872$. Comparing MRP with IPW, Figure~\ref{abcd-fig} presents the mean estimates of cognition test scores and 95\% CIs for the overall population and four sociodemographic subgroups of U.S. 9- or 10-year-old children. Detailed values are found in the Supplementary Material~\citep{mrp:supp:ys22}. For the overall mean score estimates, the MRP estimates (85.91) are slightly different from the weighted estimate (86.11), and both are lower than the sample estimate without adjustments (86.20). MRP generates higher scores for girls from single-parent families ($n=1565$, $\sim 83.17 \mbox{ vs. } 82.91 \mbox{ vs. } 82.72$), low-income white ($n=291$, $\sim 83.30 \mbox{ vs. } 82.29 \mbox{ vs. } 82.64$) and Black children ($n=1782$, $\sim 79.55 \mbox{ vs. } 79.04 \mbox{ vs. } 79.05$), but with overlapping 95\% CIs. For those from large households (more than five members) with single parents not in the labor force ($n=132$), the MRP estimates ($\sim 80.30$) are significantly lower than the weighted ($86.21$) and sample estimates ($86.43$), showing disparity. The variances of MRP estimators are lower than those of the weighted and unweighted sample estimators, illustrating efficiency gains. 

The IPW approach assumes that the ABCD survey is a quasi-random sample, and the inclusion probabilities depend on the main effects of age, sex, race/ethnicity, family income, family type, household size, patients' labor force status, and census region. Extreme weights have been trimmed. In contrast, MRP is outcome-dependent, uses the highly predictive auxiliary information of the cognitive assessments, and adjusts for differential response propensities. Under MRP, the outcome model assumes that the cognition measure follows a normal distribution, and the mean structure depends on the main effects of the first seven auxiliary variables, except for the census region. The poststratification adjustment accounts for the correlations among these seven variables and assumes that the inclusion probabilities vary across the cross-tabulation cells. The MRP estimates compose model-based predictions weighted by population cell counts and have efficiency gains. The weighted estimates use the raw values that are subject to measurement error. The poststratification of MRP borrows the joint distribution of the auxiliary variables from ACS; however, the weighting adjustment only uses their marginal distributions. It is possible that the joint distributions of the auxiliary variables are substantially different between the population and the sample, and the high-order interaction terms affect the sample inclusion propensities, leading to different estimates. The improvement is the efficiency gain from model predictions based on auxiliary information. 

The three MRP estimates are similar for the five mean estimates, with MRP-R having slightly larger variances due to the ACS sampling error. Since the effect of inclusion probabilities is small in the outcome model, MRP-INT and MRP-P generate almost identical estimates. Our application considers seven auxiliary variables, all included in the models for the outcome and inclusion probability. We do not expect inclusion probabilities to offer extra benefits. However, MRP-INT could improve MRP-P if the outcome model omits auxiliary variables accounted for in the inclusion probability model. In the Supplementary Material~\citep{mrp:supp:ys22}, we provide an additional investigation by removing two predictors, family income and labor force status, from the outcome model, but the model for inclusion probabilities keeps all seven variables.
The results show that MRP-INT with estimated inclusion probabilities as predictors outperform MRP-P for the subgroup estimates.

Our study shows cognitive performance variation across child groups with diverse sociodemographic and familial characteristics, and the use of auxiliary variables with the ACS study yields slightly different results. Literature studies show that socioeconomic disadvantage has been linked with cognitive deficits~\citep{cognitive-Bradley02}. Since the ABCD study oversamples children with high socioeconomic status, we would expect that the sample data overestimate the overall population mean of cognition test scores. We find that both MRP-adjusted or weighted values decrease, which is reasonable. Incorporating auxiliary information into MRP highlights the potentially substantial impacts of non-representative sampling on inferences. This comparison between weighting and MRP is a sensitivity analysis with different approaches yielding different conclusions, requiring further investigations for external validation. External validity assessment of different findings requires additional information and substantive knowledge, discussed below.

\section{Discussion}
\label{discussion}

Large-scale nonprobability surveys quickly emerge, demanding qualified auxiliary information and robust statistical adjustments to achieve representative inferences. To improve survey estimates for population inferences, MRP integrates nonprobability samples with population control information estimated from probability surveys, adjusting for selection/nonresponse bias and data sparsity. MRP constructs poststratification cells, fits hierarchical models, and pools cell estimates weighted by the population cell counts, with inference validity relying on the poststratification structure and model specification. MRP assumes quasi-randomization given the auxiliary variables, where the inclusion probabilities are treated as equal inside cells. MRP fits a superpopulation model to predict outcome values of nonsampled units. We have demonstrated that MRP can balance robustness and efficiency, with improvements, especially on small area estimation. 

Our simulation and application to the ABCD study show that MRP performances depend on the outcome model specification, with a similar requirement by alternative methods. Highly predictive auxiliary variables are key to success, and we recommend modeling inclusion probabilities conditional on auxiliary variables and adding a flexible function of estimated inclusion probabilities in the mean structure for better outcome model fitting performances. We present a framework for statistical data integration and robust inferences of probability and nonprobability surveys. With a small reference sample to be integrated with a nonprobability sample, the estimation of population cell sizes adds extra variability, as shown in the WFPBB implementation. 

This comprehensive study of MRP opens up several interesting future extensions. First, a sensitivity analysis framework will offer insights into potential bias if the sample inclusion is non-ignorable and still correlated with the survey outcomes conditional on the available auxiliary information. \cite{Little:west18} have proposed a measure of the degree of departure from ignorable sample selection. Depending on the magnitude of the correlation structure, a sensitivity index under MRP can be developed to reflect the range of potential bias values. Second, integrating multiple datasets, beyond two samples, can supplement the list of predictive auxiliary variables that can be sequentially imputed. Linking auxiliary information across multiple data sources has become a research priority for most statistical agencies. Data integration techniques learn about the common target population structure while accounting for the heterogeneity of different studies. Existing databases and administrative records could provide unprecedented information. Third, MRP focuses on modeling a single survey outcome and can extend to multivariate outcomes. The selection of auxiliary variables might become the union of those predictive for any outcome. The synthetic population prediction can apply state-of-the-art machine learning and deep learning algorithms to improve accuracy. Fourth, MRP brings methodological and practical challenges to model evaluation. The evaluation should include the goodness of fit in the sample data and the representation of the target population. Model evaluation in terms of generalizability is challenging in practical application studies~\citep{validity-Louis16}. Election forecasting can be checked with actual results; however, a gold standard does not exist in most areas. This is an area that needs further development and collaboration across multi-disciplines. Last, software development allowing quick implementation of the integrated MRP framework and synthetic population generation with predictions of inclusion probabilities for nonsampled cells is critical for general use.

In this application, we focus on the cognition assessment at baseline in comparison to sociodemographic groups in the ABCD study and have demonstrated the use of auxiliary information in MRP to make generalizable inferences. The ABCD study is a longitudinal study and advocates ``population neuroscience" for child development studies. Our future work also includes nonresponse bias adjustment due to attrition and population-based inferences with the brain and neurological outcome measures and methodology developments combining MRP with image modeling.

Sound design and analysis approaches are essential, emphasizing both soliciting predictive auxiliary information for analysis and improving the data collection process. Another promising direction is building an integrated database combining multiple data sources to inform design and analysis and promote open science.

\section*{Acknowledgements}
This work is funded by the National Science Foundation (SES1760133) and the National Institutes of Health (R21HD105204; U01MD017867). We thank the Editor, Associate Editor, and three referees for their thoughtful comments, Richard Valiant for sharing computational codes of simulation studies and helpful comments, and Shiro Kuriwaki for the careful reading and helpful suggestions on the manuscript formulation.

\section*{Supplementary Material}
\appendix

The supplement includes details of the simulation and application studies. We describe the model specification and implementation for the three multilevel regression and poststratification (MRP) approaches: MRP that uses all population cells for prediction with known population cell sizes (MRP-P), MRP that uses available cells in the reference probability sample for poststratification with unknown population cell sizes (MRP-R), and an integrated MRP by adding estimated inclusion probabilities as predictors in the outcome models and predicting all population cells with known population cell sizes (MRP-INT).  We present the detailed values of the outputs in tables.

\section{Simulation studies}

\subsection{Outcome model and prior specification}
\label{sim-prior}
We simulate an outcome variable $Y$ from different pre-specified models and implement two simulation scenarios: one with a correctly specified model and the other with an incorrect specification, for all the evaluated methods. In the correct model specification, the data generation model has main effects: $Y \sim age + race+ edu+inc$. For the case with model misspecification, the data generation model has all main effects and two interaction terms: $Y \sim age + race+ edu+inc+age * edu +race * inc$.

The setups of the methods are the same between the two different data generation scenarios. The outcome model in the estimation for $Y$ is the same across MRP-P, MRP-R, the general regression (GREG), and doubly robust (DR) estimators, a simple linear regression model with all main effects $Y \sim age + race+ edu+inc$. The model under MRP-INT is a multilevel regression model and has additional predictors, including the estimated inclusion probabilities $\psi_j$'s and a varying intercept across groups determined by mapping the rounded unique $\psi_j$ values with one digit. In {\em rstanarm}, the outcome model under MRP-INT is $Y \sim age + race+ edu+inc+ \psi+ (1 \mid \psi)$, where $(1 \mid \psi)$ is the varying intercept across discrete $\psi_j$ values. We include the four main effects in a logistic regression to predict response propensities in the inverse propensity score weighting (IPW) estimators. Raking adjusts for the marginal distributions of the four variables. In the incorrect model specification, all methods have failed to include the interaction terms, $(age * edu, race * inc)$. 

For all MRP outcome models, we use the default prior choices for regression models in {\em rstanarm}~\citep{rstanarm-prior}. The intercept is given a normal prior distribution with the mean as the expected value of $Y$ with the predictors set to their mean value in the data (around $-2.7$) and standard deviation (sd) of $2.5 * sd(Y)$ (around 13). The coefficients of the predictor $X$ are given a normal prior with a mean of 0 and a scale of $2.5 * sd(Y)/sd(X)$, which stables estimation but does not affect estimates when the data are even moderately informative. The default prior for the standard deviation of residuals in the linear regression is an exponential distribution with a rate of $1/sd(Y)$ (around 0.2). In the MRP-INT model, the varying intercept follows a normal distribution with a mean of 0 and an unknown standard deviation to be estimated over discrete categories based on the unique $\psi_j$ values,

For each {\em stan\_glm} or {\em stan\_glmer} model, we run two Markov chain Monte Carlo (MCMC) chains with 2000 iterations each, disregard 1000 warm-up samples each, and have achieved convergence based on the traceplots and Gelman-Rubin diagnostic measure, where the $\hat{R}$ values are around 1, for the parameters. 

To mimic the practice of survey weighting analysis, we implement the MRP-INT with two steps. We first estimate the inclusion probabilities using Stan and keep the posterior median values of 2000 samples, from two MCMC chains after disregarding 1000 warm-up iterations each, as input predictors in the following outcome modeling. Unlike a fully Bayesian procedure, the use of posterior median estimates of $\psi_j$ underestimates the overall variance. However, this is a common practice with only one set of survey weights provided for analysis (not in the replication approach), for example, with the inverse response propensity score weighting, raking, and doubly-robust estimators. 

\subsection{Poststratification cell selection and size estimate}
\label{wfpbb}
The MRP-P and MRP-INT approaches use all population cells with known counts. The MRP-R method uses available cells in the reference sample and applies the weighted finite population Bayesian bootstrap (WFPBB) to estimate the population cell counts. \cite{fpbb14} propose WFPBB as a nonparametric method of generating synthetic populations that can be analyzed as simple random samples by ``undoing" the complex sampling design and accounting for weights. The idea is to draw from the posterior predictive distribution of non-observed (nob) data given the observed (obs) data and weights. In our simulation study, the reference sample is a simple random sample with equal base weights $w_i=N/n=50000/100$. We embed the WFPBB in the MRP-R implementation following the steps below. 

\begin{enumerate}
    \item \textbf{Resample via Bayesian bootstrap (BB)}: To capture the sampling variability of drawing from the posterior distribution of the population parameters given the data from the ``parent" (original) sample, we generate \textit{L=100} number of BB samples: $B_1,\ldots, B_L$, each of size $n=1000$.
    \item \textbf{Recalibrate weights}: For each BB sample $B_l$, we recalibrate the bootstrap weights by multiplying the base weights by the number of replicates for unit $i$ in $B_l$ ($r^{l}_{i}$) and normalizing the weights to sum to the population size $N$, so that $w^l_i=N\frac{w_ir^{l}_{i}}{\sum_j w_jr^l_j}$.
    \item \textbf{Use WFPBB to incorporate sampling weights}:  Construct the initial P\'{o}lya urn based on the BB sample with their corresponding replicate weights $w^l_i$'s and draw $N-n$ units with probability
    \begin{equation}
       \frac{w^l_i - 1 + l_{i, k-1}(N-n)/n}{N-n + (k-1)(N-n)/n},
        \label{eqpolya}
     \end{equation}
for the $k$th draw, $k \in \{1, \ldots, (N-n)\}$, where $l_{i, k-1}$ is the number of bootstrap selections of $(z_i,x_i)$ among the elements present in our urn at the $k-1$ draw. The draws form the WFPBB sample $S_{l}$ of size $N$. 
  
\item \textbf{Bayesian inference}: For each of the samples $S_{l} \in \{S_{1},\ldots,S_{l}\}$ we obtain estimates $\hat{N}_j^{(l)}$'s. 
\end{enumerate}

The WFPBB approach generates 100 synthetic populations with the same size $N=50,000$, i.e., 100 posterior samples of $\hat{N}_j$'s for poststratification under MRP-R. 

\subsection{Detailed outputs}

In Tables~\ref{cor} and \ref{incor} we present the estimates for bias, standard error (SE), root mean squared error (RMSE) and coverage rates of the 95\% confidence intervals under the two simulation scenarios for the seven methods (DR: doubly robust estimator, IPW: inverse propensity weighting estimator, GREG: generalized regression estimator, MRP-INT: an integrated MRP by adding estimated inclusion probabilities as predictors in the outcome models and predicting all population cells with known population cell sizes, MRP-P: MRP that uses all population cells for poststratification with known population cell sizes, and MRP-R: MRP that uses available cells in the reference probability sample for poststratification with unknown population cell sizes).

\begin{table}[h]
\centering
\footnotesize
\caption{\em \footnotesize Outputs with correct outcome models and adjustment variables.}
\label{cor}
\vspace{-0.25cm}
\begin{tabular}{llrrrr}
  \hline
 Method & Group & Bias & RMSE & SE & Coverage rates \\ 
  \hline
MRP-P & AGE:18-24 & -0.00 & 0.09 & 0.09 & 0.94 \\ 
MRP-P & AGE:25-34 & -0.00 & 0.06 & 0.06 & 0.95 \\ 
MRP-P & AGE:35-44 & -0.00 & 0.07 & 0.07 & 0.97 \\ 
MRP-P & AGE:45-54 & -0.01 & 0.07 & 0.07 & 0.95 \\ 
MRP-P & AGE:55-64 & 0.00 & 0.09 & 0.09 & 0.92 \\ 
MRP-P & AGE:65+ & -0.01 & 0.14 & 0.14 & 0.96 \\ 
MRP-P & Overall & -0.00 & 0.04 & 0.04 & 0.94 \\ 
MRP-R & AGE:18-24 & -0.01 & 0.36 & 0.36 & 0.94 \\ 
MRP-R & AGE:25-34 & -0.00 & 0.27 & 0.27 & 0.96 \\ 
MRP-R & AGE:35-44 & -0.00 & 0.22 & 0.23 & 0.94 \\ 
MRP-R & AGE:45-54 & 0.00 & 0.22 & 0.22 & 0.95 \\ 
MRP-R & AGE:55-64 & -0.00 & 0.24 & 0.24 & 0.94 \\ 
 MRP-R & AGE:65+ & -0.00 & 0.22 & 0.17 & 0.84 \\ 
MRP-R & Overall & -0.00 & 0.13 & 0.12 & 0.91 \\ 
MRP-INT & AGE:18-24 & -0.00 & 0.09 & 0.09 & 0.94 \\ 
MRP-INT & AGE:25-34 & -0.00 & 0.06 & 0.06 & 0.96 \\ 
MRP-INT & AGE:35-44 & -0.00 & 0.08 & 0.07 & 0.96 \\ 
MRP-INT & AGE:45-54 & -0.01 & 0.07 & 0.07 & 0.95 \\ 
MRP-INT & AGE:55-64 & 0.00 & 0.09 & 0.09 & 0.95 \\ 
 MRP-INT & AGE:65+ & 0.01 & 0.14 & 0.15 & 0.96 \\ 
MRP-INT & Overall & -0.00 & 0.04 & 0.04 & 0.95 \\ 
IPW & AGE:18-24 & -0.01 & 0.26 & 0.26 & 0.96 \\ 
IPW & AGE:25-34 & -0.01 & 0.20 & 0.18 & 0.91 \\ 
 IPW & AGE:35-44 & -0.11 & 0.31 & 0.24 & 0.89 \\ 
IPW & AGE:45-54 & -0.04 & 0.28 & 0.24 & 0.92 \\ 
 IPW & AGE:55-64 & 0.01 & 0.29 & 0.28 & 0.93 \\ 
 IPW & AGE:65+ & 0.15 & 0.52 & 0.40 & 0.86 \\ 
 IPW & Overall & -0.01 & 0.16 & 0.17 & 0.95 \\ 
GREG & AGE:18-24 & 0.13 & 0.53 & 0.37 & 0.86 \\ 
GREG & AGE:25-34 & 0.15 & 0.67 & 0.32 & 0.59 \\ 
 GREG & AGE:35-44 & -0.11 & 0.29 & 0.23 & 0.84 \\ 
 GREG & AGE:45-54 & -0.05 & 0.25 & 0.22 & 0.88 \\ 
 GREG & AGE:55-64 & -0.00 & 0.27 & 0.27 & 0.92 \\ 
GREG & AGE:65+ & 0.23 & 0.64 & 0.39 & 0.78 \\ 
GREG & Overall & -0.00 & 0.04 & 0.04 & 0.94 \\ 
Raking & AGE:18-24 & 0.03 & 0.28 & 0.37 & 0.94 \\ 
Raking & AGE:25-34 & 0.02 & 0.21 & 0.31 & 0.91 \\ 
Raking & AGE:35-44 & -0.14 & 0.32 & 0.23 & 0.80 \\ 
Raking & AGE:45-54 & -0.02 & 0.22 & 0.23 & 0.92 \\ 
Raking & AGE:55-64 & 0.04 & 0.29 & 0.28 & 0.91 \\ 
Raking & AGE:65+ & 0.15 & 0.46 & 0.38 & 0.90 \\ 
Raking & Overall & 0.00 & 0.05 & 0.05 & 0.94 \\ 
DR & AGE:18-24 & 0.02 & 0.27 & 0.26 & 0.96 \\ 
DR & AGE:25-34 & 0.01 & 0.20 & 0.18 & 0.90 \\ 
DR & AGE:35-44 & -0.13 & 0.31 & 0.24 & 0.79 \\ 
DR & AGE:45-54 & -0.02 & 0.22 & 0.24 & 0.93 \\ 
DR & AGE:55-64 & 0.04 & 0.29 & 0.28 & 0.91 \\ 
DR & AGE:65+ & 0.12 & 0.44 & 0.39 & 0.88 \\ 
DR & Overall & -0.00 & 0.05 & 0.17 & 0.94 \\ 
   \hline
\end{tabular}
\end{table}

\begin{table}[h]
\centering
\footnotesize
\caption{\em \footnotesize Outputs with incorrect outcome models and adjustment variables.}
\label{incor}
\vspace{-0.25cm}
\begin{tabular}{llrrrr}
  \hline
 Method & Group & Bias & RMSE & SE & Coverage rates \\ 
  \hline
MRP-P & AGE:18-24 & -0.04 & 0.17 & 0.15 & 0.92 \\ 
MRP-P & AGE:25-34 & 0.02 & 0.14 & 0.10 & 0.78 \\ 
MRP-P & AGE:35-44 & 0.02 & 0.15 & 0.12 & 0.88 \\ 
MRP-P & AGE:45-54 & 0.02 & 0.16 & 0.12 & 0.89 \\ 
 MRP-P & AGE:55-64 & 0.26 & 0.19 & 0.15 & 0.87 \\ 
MRP-P & AGE:65+ & -1.62 & 0.96 & 0.24 & 0.13 \\ 
MRP-P & Overall & -0.11 & 0.21 & 0.07 & 0.33 \\ 
MRP-R & AGE:18-24 & -0.04 & 0.40 & 0.35 & 0.91 \\ 
MRP-R & AGE:25-34 & 0.02 & 0.29 & 0.26 & 0.91 \\ 
MRP-R & AGE:35-44 & 0.02 & 0.24 & 0.21 & 0.89 \\ 
MRP-R & AGE:45-54 & 0.03 & 0.29 & 0.20 & 0.81 \\ 
MRP-R & AGE:55-64 & 0.18 & 0.28 & 0.22 & 0.88 \\ 
MRP-R & AGE:65+ & -1.53 & 0.95 & 0.17 & 0.10 \\ 
MRP-R & Overall & -0.09 & 0.24 & 0.15 & 0.72 \\ 
MRP-INT & AGE:18-24 & -0.03 & 0.17 & 0.14 & 0.90 \\ 
MRP-INT & AGE:25-34 & 0.01 & 0.12 & 0.09 & 0.85 \\ 
MRP-INT & AGE:35-44 & 0.02 & 0.13 & 0.11 & 0.89 \\ 
MRP-INT & AGE:45-54 & 0.03 & 0.21 & 0.11 & 0.61 \\ 
MRP-INT & AGE:55-64 & 0.15 & 0.15 & 0.13 & 0.92 \\ 
MRP-INT & AGE:65+ & -0.46 & 0.43 & 0.26 & 0.81 \\ 
MRP-INT & Overall & -0.02 & 0.10 & 0.07 & 0.88 \\ 
IPW & AGE:18-24 & 0.02 & 0.27 & 0.28 & 0.96 \\ 
IPW & AGE:25-34 & -0.01 & 0.22 & 0.21 & 0.91 \\ 
IPW & AGE:35-44 & -0.04 & 0.28 & 0.25 & 0.91 \\ 
IPW & AGE:45-54 & -0.00 & 0.26 & 0.25 & 0.93 \\ 
IPW & AGE:55-64 & -0.08 & 0.31 & 0.31 & 0.93 \\ 
IPW & AGE:65+ & -1.34 & 0.93 & 0.52 & 0.71 \\ 
IPW & Overall & -0.07 & 0.24 & 0.19 & 0.90 \\ 
GREG & AGE:18-24 & 0.18 & 0.64 & 0.40 & 0.80 \\ 
GREG & AGE:25-34 & -0.03 & 0.41 & 0.37 & 0.90 \\ 
GREG & AGE:35-44 & -0.03 & 0.26 & 0.24 & 0.87 \\ 
GREG & AGE:45-54 & -0.01 & 0.24 & 0.23 & 0.95 \\ 
 GREG & AGE:55-64 & -0.07 & 0.30 & 0.29 & 0.92 \\ 
GREG & AGE:65+ & -1.78 & 1.09 & 0.49 & 0.53 \\ 
GREG & Overall & -0.11 & 0.22 & 0.10 & 0.47 \\ 
Raking & AGE:18-24 & 0.07 & 0.34 & 0.40 & 0.92 \\ 
 Raking & AGE:25-34 & -0.02 & 0.24 & 0.35 & 0.89 \\ 
Raking & AGE:35-44 & -0.05 & 0.28 & 0.24 & 0.85 \\ 
Raking & AGE:45-54 & -0.00 & 0.23 & 0.23 & 0.95 \\ 
Raking & AGE:55-64 & -0.28 & 0.34 & 0.31 & 0.91 \\ 
 Raking & AGE:65+ & -1.12 & 0.76 & 0.52 & 0.77 \\ 
Raking & Overall & -0.05 & 0.13 & 0.11 & 0.83 \\ 
DR & AGE:18-24 & 0.06 & 0.33 & 0.27 & 0.92 \\ 
DR & AGE:25-34 & -0.02 & 0.25 & 0.21 & 0.90 \\ 
DR & AGE:35-44 & -0.05 & 0.28 & 0.25 & 0.86 \\ 
DR & AGE:45-54 & -0.00 & 0.23 & 0.25 & 0.94 \\ 
DR & AGE:55-64 & -0.26 & 0.31 & 0.31 & 0.94 \\ 
 DR & AGE:65+ & -1.08 & 0.75 & 0.53 & 0.78 \\ 
DR & Overall & -0.05 & 0.13 & 0.19 & 0.82 \\ 
   \hline
\end{tabular}
\end{table}

\section{Application to the Adolescent Brain Cognitive Development Study}

We apply MRP-P, MRP-R, and MRP-INT to the Adolescent Brain Cognitive Development (ABCD) study and compare with IPW and unweighted sample mean estimators. The outcome model under MRP-P and MRP-R for the cognitive test score is a linear regression: $Y \sim age + sex + race/ethnicity + family income + family type + household size + labor force status$, with random effects introduced to the coefficients of multiple-category predictors: race/ethnicity, income, household size, and parents' labor force status. All seven variables are used to estimate the inclusion probabilities $\psi_j$. The model under MRP-INT has additional predictors, including the estimated inclusion probabilities $\psi_j$'s and a varying intercept across groups determined by mapping the rounded unique $\psi_j$ values with one digit. In {\em rstanarm}, the outcome model under MRP-INT is $Y \sim age + sex + (1 \mid race/ethnicity) + (1 \mid family income) + family type + (1 \mid household size) + (1 \mid labor force status)+ \psi+ (1 \mid \psi)$, where $(1 \mid \psi)$ is the varying intercept across discrete $\psi_j$ values. For all MRP outcome models, we use the default prior choices for regression models in {\em rstanarm}, which are the same as the setting above in Section~\ref{sim-prior}. 

In MRP-R, to account for the sampling error of the American Community Survey (ACS), we implement the WFPBB algorithm, described in Section~\ref{wfpbb}, and generate $L=10$ synthetic populations, each with a size of $N=8,211,605$ as the sum of ACS weights. We apply the combining rules~\citep{fpbb14}, rather than using the Bayesian posterior sample inference, to obtain the variances of the population mean estimators. 

With seven auxiliary variables, the detailed values of the outputs comparing MRP-P, MRP-R, MRP-INT, IPW, and unweighted sample mean estimators are presented in Table~\ref{abcd-est}.

\begin{table}[ht]
\center
\footnotesize
\caption{Finite population inferences of average cognition test scores by groups with seven auxiliary variables (95\% confidence intervals in parenthesis).}
\label{abcd-est}
\begin{tabular}{llllll}
\hline
Quantity&MRP-INT&MRP-P&MRP&IPW&Sample\\
\hline
Overall &85.91 &85.91 & 85.92 &86.11 &86.20 \\
(n=11872) & (85.75, 86.07) &(85.75, 86.07) &(85.73, 86.10)&(85.93, 86.30)&(86.03, 86.36)\\
\hline
Single-parent girls& 83.17 &83.15&83.16 &82.91&82.72\\
(n=1565) &(82.85, 83.49) &(82.83, 83.47)& (82.78, 83.54)&  (82.43, 83.4) & (82.28, 83.17)\\
\hline
Black  &79.55 &79.57 &79.56 &79.04 &79.05 \\
(n=1782)&(79.18, 79.93)&(79.21, 79.92) &(79.14, 79.97) &(78.57, 79.51)&(78.64, 79.47)\\
\hline
Low-income ($<25$K)&83.30 &83.24&83.25 &82.29&82.64\\
white (n=291)&(82.84, 83.77)& (82.78, 83.68)&(82.73, 83.77)&(81.13, 83.45)&(81.6, 83.67)\\
\hline
Single parent, not in LF&80.24&80.29&80.30&86.21&86.43\\
HH size $> 5$ (n=132)&(79.57, 80.92)&(79.63, 80.98)&(79.50, 81.10)&(85.67, 86.75)&(85.94, 86.91)\\
\hline
\end{tabular}
\\
\scriptsize{\em IPW: inverse propensity score weighting, MRP: multilevel regression and poststratification, MRP-INT: an integrated MRP by adding estimated inclusion probabilities as predictors in the outcome models and predicting all population cells with known population cell sizes, MRP-P: MRP that uses all population cells with known population cell sizes, and MRP-R: MRP that uses available cells in the reference probability sample with unknown population cell sizes.}
\end{table}

Table~\ref{abcd-est} shows that the three MRP estimators are similar. This is expected as all seven auxiliary variables are included in the outcome models, where the estimated inclusion probabilities will not offer additional prediction power. To demonstrate the potential improvement of MRP-INT over MRP-P, we consider a different outcome model specification in the ABCD study. Since family income and labor force status are predictive, we remove them from the outcome model and ``assume" a misspecified model.

The outcome model under MRP-P and MRP-R now has five predictors: $Y \sim age + sex + race/ethnicity +  family type + household size $, with random effects introduced to the coefficients of multiple-category predictors: race/ethnicity and household size. However, all seven variables are used to estimate the inclusion probabilities $\psi_j$. The model under MRP-INT is $Y \sim age + sex + (1 \mid race/ethnicity) + family type + (1 \mid household size) + \psi+ (1 \mid \psi)$, where $(1 \mid \psi)$ is the varying intercept across discrete $\psi_j$ values. All other settings are the same as above. 

\begin{figure}[ht]
\begin{tabular}{c}
\includegraphics[width=0.95\textwidth]{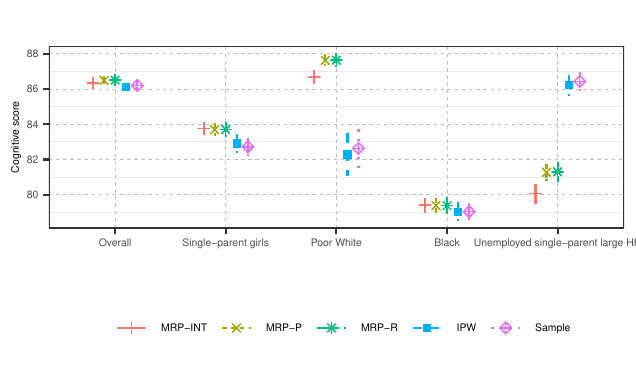}
\end{tabular}
\caption{\em \small Finite population inferences of average cognition test scores by groups. {\bf The outcome model has five auxiliary variables.}The error bars are 95\% confidence intervals. HH: household, IPW: inverse propensity weighting estimator, MRP: multilevel regression and poststratification, MRP-INT: an integrated MRP by adding estimated inclusion probabilities as predictors in the outcome models and predicting all population cells with known population cell sizes, MRP-P: MRP that uses all population cells with known population cell sizes, and MRP-R: MRP that uses available cells in the reference probability sample with unknown population cell sizes. }
\label{abcd-fig-new}
\end{figure}

Figure~\ref{abcd-fig-new} presents the new outputs with five auxiliary variables. We can see that MRP-INT estimates are different from those of MRP-P for the poor white and children from large households (more than five members) with single parents not in the labor force. The MRP-P and MRP-R estimates for these two subgroups are substantially different from those when the outcome model includes family income and labor force status, showing potentially biased estimates. However, the MRP-INT estimate for the subgroup of children from large households (more than five members) with single parents not in the labor force is similar before and after removing the two predictors. The MRP-INT estimate for poor white children has smaller differences than those of MRP-P estimates. Overall, MRP-INT improves MRP-P estimates and offers protection against model misspecification. The improvement depends on how much information the estimated inclusion probabilities bring to increase the prediction power, since the model under MRP-INT is not the actual specification.

\bibliography{ys-2023b.bib}
\bibliographystyle{chicago}

\end{document}